\documentclass[rn,11pt]{ucl_document} 
\usepackage{graphicx}
\usepackage{hyperref} 

\newlength{\halfwidth}
\begin{document}

\title{Fast Generation of Big Random Binary Trees}

\author{W. B. Langdon}

\date{13 January 2020}

\documentnumber{20/01}

\maketitle

\begin{abstract}
random\_tree() is a linear time and space C++ implementation
able
to create
trees 
of up to a billion nodes
for genetic programming
and genetic improvement experiments.
A 3.60GHz CPU
can generate
more than 18~million random nodes 
for GP program trees per second.

\end{abstract}

\section{Introduction}

In most cases
genetic programming (GP)
represents programs as trees
\cite{koza:book,poli08:fieldguide}.
GP
has always needed pseudo random trees.
Firstly for the initial random population
and secondly
typically mutation operations replace a small part of an existing program
with
a small randomly generated subtree.
In many cases the same algorithm,
albeit with different parameters,
is used both to create the initial population 
and to mutate subtrees.
Koza's~\cite{koza:book}
ramped half-and-half is often used for both.
Although ramped half-and-half creates trees of diverse sizes and
shapes it does not sample the space of trees uniformly.
(For example, it samples bushy nearly full trees more heavily.
Notice that solutions to some problems,
such as the parity problems,
are denser in this bushy sub-space than in general~\cite{langdon:1999:sptfs}.)

Bohm~\cite{bohm:1996:eui}
and
Iba
~\cite{iba:1996:rtgGP}
proposed sampling the space of programs more uniformly by settling on
a tree size and then randomly sampling trees of exactly that size
uniformly at random.
Following Iba~\cite{iba:1995:rtgTR}'s linear, O(tree size), algorithm,
I implemented in Andy Singleton's C++ GPquick~\cite{singleton:byte}
a linear algorithm which first chose at random from a range of tree
sizes and then generated uniformly at random a tree of the chosen
size~\cite{langdon:2000:fairxo}.
Like ramped half-and-half,
this was used to generate both the initial population and to define
another subtree mutation operator.
Note initial random programs were small (perhaps 50 nodes),
indeed subtrees required for mutation are often no more than five nodes.
The C++ implementation allowed GP program trees containing functions
with up to four arguments.

Before genetic programming is started, 
the functions and the number of their arguments
(their arities)
must be defined.
(This is known as the functionset~\cite{koza:book,poli08:fieldguide}).
GPquick allows this to be done via an external file,
\verb'prim.dat'.
After \verb'prim.dat' had been read,
my rand\_tree extension to GPquick,
created various tables containing the number of trees with
each legal combination of arities
\cite{segdewick:1996:aa}.
To avoid repeatedly calculating the tree counts,
these tables are consulted during GP run time.
When generating a new random tree,
which type of tree
(i.e.\ which combination of arities)
is chosen uniformly at random based on the number of programs 
of the required size.
Although linear in tree size,
the implementation was not desperately efficient,
and criticised as such~\cite{luke:2000:2ftcaGP},
but it was felt not to be too important
since the random tree were small
and GP runtime is typically 
dominated by fitness evaluation rather than genetic operations such
as subtree mutation.
That was the situation for twenty years.

With the availability of fast parallel hardware
\cite{langdon:2019:gpquick}
GPquick 
has been used in greatly extended evolutionary runs
of a hundred of thousand
\cite{langdon:RN1705,Langdon:2017:GECCO},
even a million generations
\cite{langdon:RN1901,%
Langdon:2019:alife}.
Naturally, with static fitness functions and no constraints,
enormous trees
(hundreds of millions of nodes)
were evolved.
These experiments took weeks or months to run.
(GPquick allows terminals and functions to have side effects
and does not exploit their absence
\cite{Handley:1994:DAGpcp,%
mcphee:1998:sutherland}.)

I hope to significantly reduce run time by evolving considerable
further improvements in GPquick.
The question of how to test it on big trees has been address
by replacing the evolved trees 
with randomly generated trees.
The huge evolved trees resemble random trees
\cite{langdon:RN1705,%
Langdon:2019:alife}.
The rand\_tree C++ implementation (described above) is
hopelessly inefficient for trees of a billion nodes,
therefore
it was rewritten.
The new RAND\_TREE2\_FASTER C++ code
deals only with binary trees.
The algorithm is still 
Iba's~\cite{iba:1995:rtgTR}
and, if need be,
could be extended to deal with trees 
containing functions with one argument
and/or functions with more than two arguments.

\section{Converting Random Permutations into Random Binary Trees}

The new random\_tree() 
starts by generating a deterministic list of
alternate $n+1$ leafs and $n$ functions
and then uses 
Knuth's shuffle
to randomise it.
That is,
starting at the beginning of the list of $2n+1$ items,
swap the current item with another item later in the list chosen at random.
Move one item at a time along the list to its end,
so that the whole list is now in a random order.
Next
random\_tree() converts this random permutation into
a random binary tree.

\begin{figure}
\centerline{%
\includegraphics[scale=0.7]{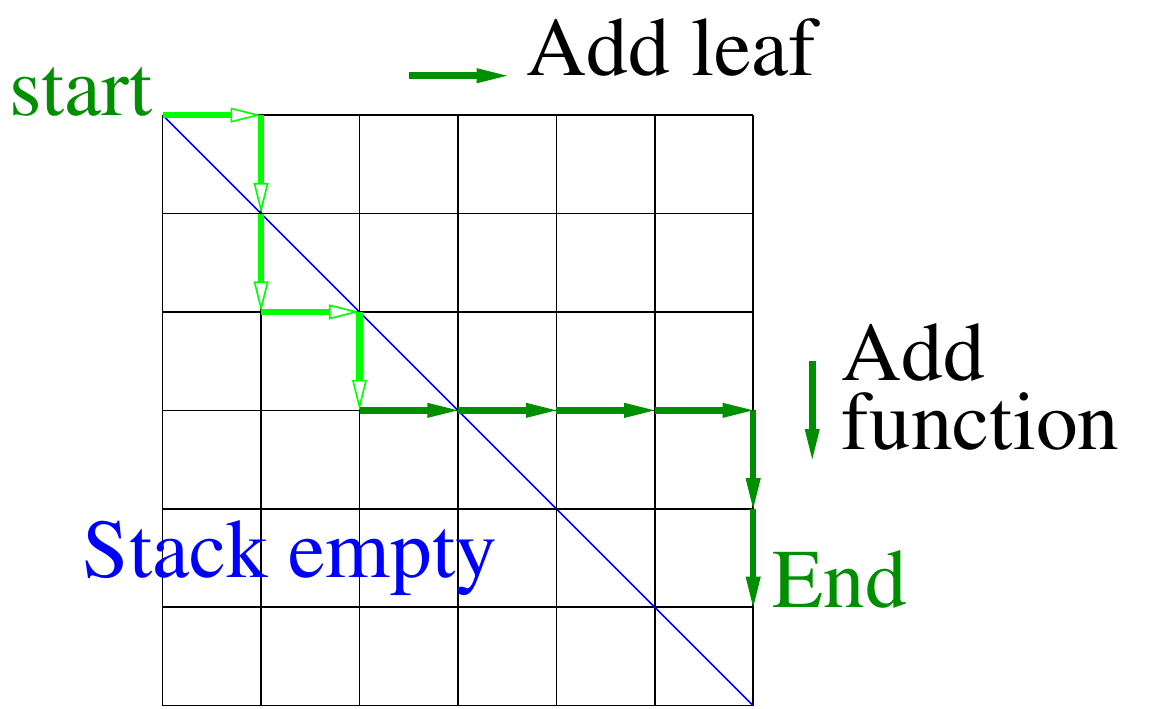} 
\includegraphics[scale=0.7]{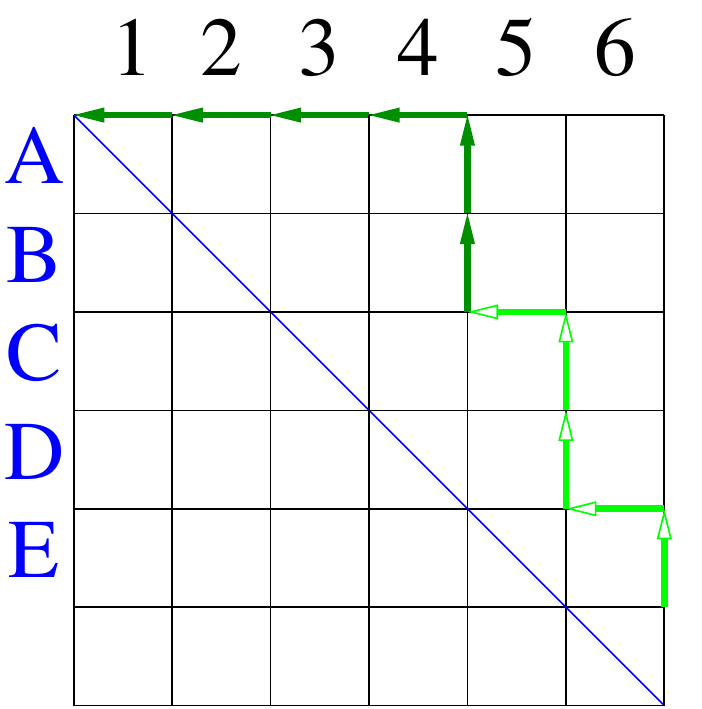} 
\includegraphics[scale=0.7]{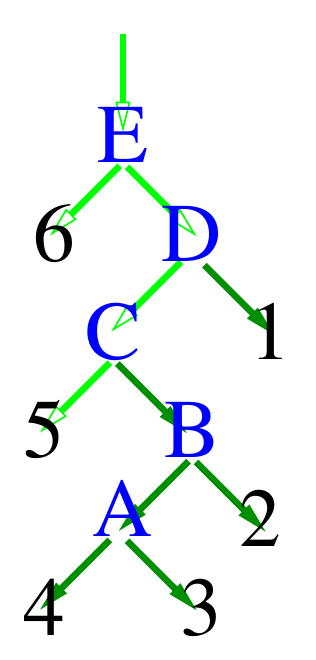}} 
\caption{\label{fig:square6}
Any combination of $n$ vertical (down) moves and
$n+1$ horizontal (right) moves
from start will reach end
without leaving the square box.
Middle: same random sequence but rotated 
to form valid binary tree,
i.e.\
stack is never
less than empty.
Right: tree.
}
\end{figure}

\pagebreak[4]
Consider a square lattice grid of side $n+1$.
If we will have a binary tree with $n$ internal nodes
(and $n+1$ leafs)
it can be placed into the square
as shown in Figure~\ref{fig:square6} (middle).
However our initial
random permutation of $n+1$ leafs and
$n$ functions 
is unlikely to be a valid tree
as it is likely to enter the half of the square
corresponding to 
having more leafs than positions to attach them.
(I.e.\ doing more stack pops than the stack has data on it.)
See also \cite[Fig~5.12 p269]{segdewick:1996:aa}.
However in O(n) steps we can find the deepest point in the 
lattice
curve.
I.e.\ the knee furthest from the blue diagonal in 
Figure~\ref{fig:square6} (left)
and remap our random list to start here.
This gives us a new lattice curve 
Figure~\ref{fig:square6} (right)
which is certain never to cross below the diagonal.
This gives the shape of the tree.

A corresponding random program is created in O(n) steps.
For each element of the (re-ordered) random list,
the GPquick SETNODE macro is used
to set the corresponding node in the GP program.
If the list item is a leaf: a leaf is chosen at random.
Otherwise, one of the binary GP functions is chosen 
at random.

\section{Performance}

The new implementation,
int~random\_tree()
rand\_tree.cc Revision:~1.43, 
is not of the utmost efficiency.
For example,
for ease of debugging and modularity,
separate passes are used to create
the random tree and labelled it
(with leafs and functions
in order to convert it into a GP program).
These two passes could be combined.
This might be beneficial, 
especially for trees 
which are too big to fit into cache.

Similarly the wellformed() debug check could be disabled
and max\_depth could be calculated from the lattice
(cf.\ Figure~\ref{fig:square6})
rather than via recurse().

\begin{figure}
\centerline{\includegraphics{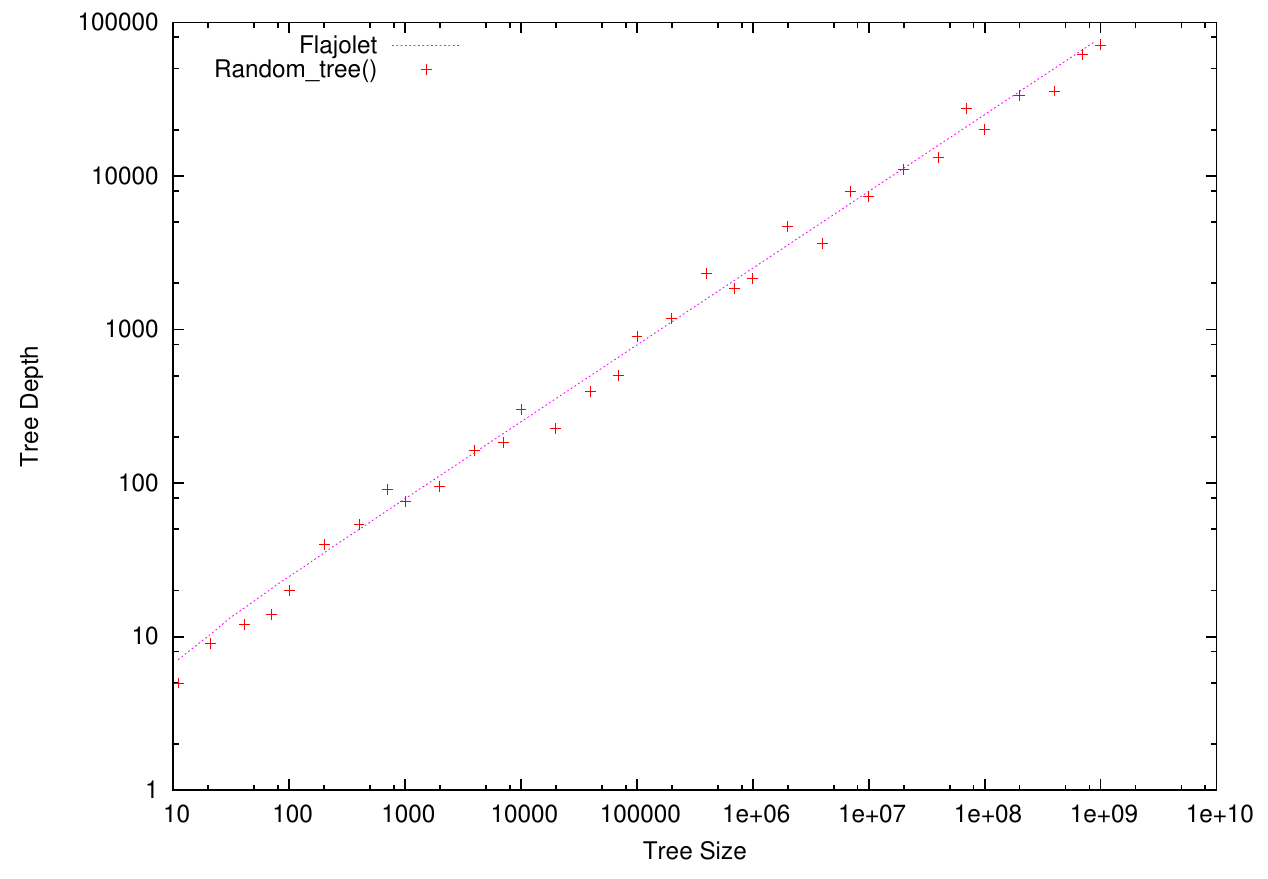}} 
\vspace*{-2ex}
\caption{\label{fig:rand_tree}
Examples of large random binary trees.
As expected they lie near the Flajolet large tree limit
(depth $\approx\!\!\sqrt{2\pi|{\rm size}|}$).
Note log-log scales.
}
\end{figure}

\begin{figure}
\centerline{\includegraphics{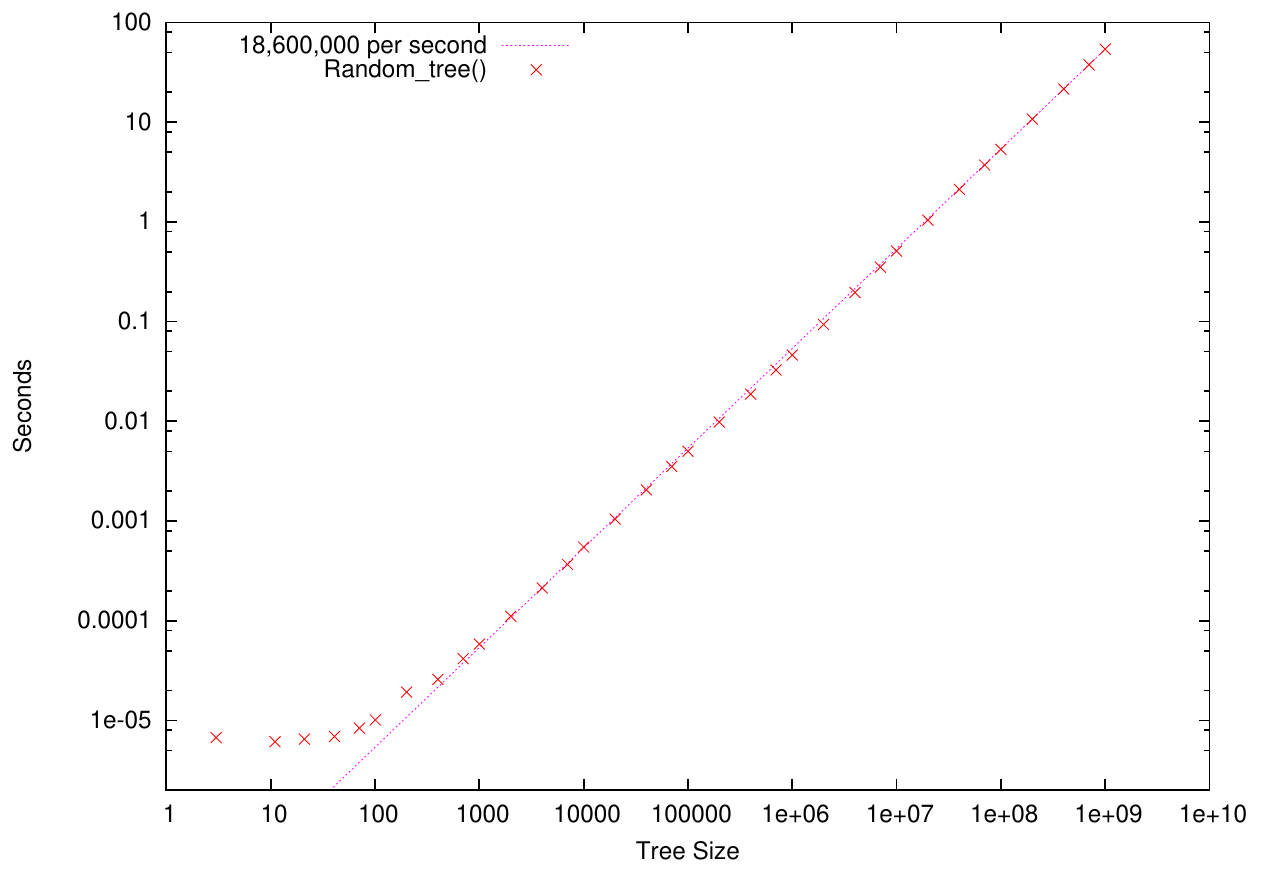}} 
\vspace*{-2ex}
\caption{\label{fig:time_rand}
Time taken to create random binary trees
on one core of Intel i7-4790 3.60GHz desktop.
As expected run time scales linearly with tree size.
(Note log-log scales.)
}
\end{figure}

\section{Depth of Random Binary Trees}

The average height of a random binary tree
with $N$ (internal) nodes
is 
$2\sqrt{\pi N} + O(N^{1/4+\epsilon})$ for any $\epsilon>0$
\cite[page~256]{segdewick:1996:aa}.
GPquick commonly uses the size of the tree ($2N+1$).
Therefore the tree depth of a typical large random tree
is
$\approx \sqrt{2 \pi |{\rm size}|}$.
This approximation is better than 2\% accuracy for trees 
above 32\,000~\cite[page~200]{flajlet:1982:ahbt}.
Flajolet and Oldyzko
give 
not just the mean but also the limits for 
the variance and all the higher order
moments
for the distribution of random binary tree depth
in terms of gamma and Riemann Zeta functions.
They say for trees of about 20\,000 nodes,
their estimates are with 10\% of the actual values
\cite[page~210]{flajlet:1982:ahbt}.
For large trees,
they 
say the distribution of binary tree sizes of a fixed height is
Gaussian
\cite[page~212]{flajlet:1982:ahbt}.

Although it is not necessary
(see \cite{langdon:2008:eurogp,langdon:2008:SC})
GPquick,
like most GP systems,
interprets the evolved programs recursively.
The maximum depth of the recursive calls is the depth of the tree
used to store the program.
random\_tree()
returns the actual depth of the newly generated tree.
This can either be used as a sanity check that the tree can be
evaluated within the available data structures
or to set the size of those data structures.

\section{Limitations}

The new implementation is limited at various points in GPquick by the
use of 32~bit integers to describe the size of programs
and so suffers from segmentation errors (SEG\_FAULT)
when trees in the region of 2\,000\,000\,000 are requested.

I have used the short hand ``random'' to actually mean pseudo random (PRNG).
GPquick uses the Park-Miller random number generator
\cite{langdon:2008:CIGPU,Park:1988:RNG}.
The number of possible programs is vastly more than the number
of PRNG seeds
and so any real program can only be drawn from a tiny subset
of the total.
Nonetheless the programs generated appear to be sufficiently random for typical
GP use.

\vspace{-1ex}
\section*{Acknowledgements}
This work was inspired by conversations
at Dagstuhl 
\href{https://www.dagstuhl.de/en/program/calendar/semhp/?semnr=18052}
{Seminar 18052} on
Genetic Improvement of Software.

Funded by EPSRC grant
\href{http://gow.epsrc.ukri.org/NGBOViewGrant.aspx?GrantRef=EP/M025853/1}
{EP/M025853/1}.

random\_tree code (intended to be used as part of GPquick)
can be found via
\url{http://www.cs.ucl.ac.uk/staff/W.Langdon/ftp/gp-code/rand_tree.cc_r1.43}

\vspace{-2ex}
\bibliographystyle{splncs03}
\bibliography{/tmp/gp-bibliography,/tmp/references}

\begin{thebibliography}{10}
\providecommand{\url}[1]{\texttt{#1}}
\providecommand{\urlprefix}{URL }

\bibitem{koza:book}
Koza, J.R.: Genetic Programming: On the Programming of Computers by Means of
  Natural Selection. MIT Press, Cambridge, MA, USA (1992),
  \url{http://mitpress.mit.edu/books/genetic-programming}

\bibitem{poli08:fieldguide}
Poli, R., Langdon, W.B., McPhee, N.F.: A field guide to genetic programming.
  Published via \texttt{http://lulu.com} and freely available at
  \texttt{http://www.gp-field-guide.org.uk} (2008),
  \url{http://www.gp-field-guide.org.uk}, (With contributions by J. R. Koza)

\bibitem{langdon:1999:sptfs}
Langdon, W.B.: Scaling of program tree fitness spaces. Evolutionary Computation
   7(4),  399--428 (Winter 1999),
  \url{http://dx.doi.org/10.1162/evco.1999.7.4.399}

\bibitem{bohm:1996:eui}
Bohm, W., Geyer-Schulz, A.: Exact uniform initialization for genetic
  programming. In: Belew, R.K., Vose, M. (eds.) Foundations of Genetic
  Algorithms IV. pp. 379--407. Morgan Kaufmann, University of San Diego, CA,
  USA (3--5 Aug 1996),
  \url{http://cseweb.ucsd.edu/~rik/foga4/Abstracts/07-wb-abs.html}

\bibitem{iba:1996:rtgGP}
Iba, H.: Random tree generation for genetic programming. In: Voigt, H.M.,
  Ebeling, W., Rechenberg, I., Schwefel, H.P. (eds.) Parallel Problem Solving
  from Nature IV, Proceedings of the International Conference on Evolutionary
  Computation. LNCS, vol. 1141, pp. 144--153. Springer Verlag, Berlin, Germany
  (22-26 Sep 1996), \url{http://dx.doi.org/10.1007/3-540-61723-X_978}

\bibitem{iba:1995:rtgTR}
Iba, H.: Random tree generation for genetic programming. Tech. Rep.
  ETL-TR-95-35, ElectroTechnical Laboratory (ETL), 1-1-4 Umezono, Tsukuba-city,
  Ibaraki, 305, Japan (14 Nov 1995),
  \url{http://www.cs.ucl.ac.uk/staff/W.Langdon/ftp/papers/iba_1995_rtgTR.pdf}

\bibitem{singleton:byte}
Singleton, A.: Genetic programming with {C}++. BYTE pp. 171--176 (Feb 1994),
  \url{http://www.assembla.com/wiki/show/andysgp/GPQuick_Article}

\bibitem{langdon:2000:fairxo}
Langdon, W.B.: Size fair and homologous tree genetic programming crossovers.
  Genetic Programming and Evolvable Machines  1(1/2),  95--119 (Apr 2000),
  \url{http://dx.doi.org/10.1023/A:1010024515191}

\bibitem{segdewick:1996:aa}
Sedgewick, R., Flajolet, P.: An Introduction to the Analysis of Algorithms.
  Addison-Wesley (1996)

\bibitem{luke:2000:2ftcaGP}
Luke, S.: Two fast tree-creation algorithms for genetic programming. IEEE
  Transactions on Evolutionary Computation  4(3),  274--283 (Sep 2000),
  \url{http://dx.doi.org/10.1109/4235.873237}

\bibitem{langdon:2019:gpquick}
Langdon, W.B.: Parallel {GPQUICK}. In: Doerr, C. (ed.) GECCO '19: Proceedings
  of the Genetic and Evolutionary Computation Conference Companion. pp. 63--64.
  ACM, Prague, Czech Republic (Jul 13-17 2019),
  \url{http://dx.doi.org/10.1145/3319619.3326770}

\bibitem{langdon:RN1705}
Langdon, W.B.: Long-term evolution of genetic programming populations. Tech.
  Rep. RN/17/05, University College, London, London, UK (24 Mar 2017),
  \url{http://www.cs.ucl.ac.uk/fileadmin/UCL-CS/research/Research_Notes/RN_17_05.pdf},
  also available as arXiv 1843365

\bibitem{Langdon:2017:GECCO}
Langdon, W.B.: Long-term evolution of genetic programming populations. In:
  GECCO 2017: The Genetic and Evolutionary Computation Conference. pp.
  235--236. ACM, Berlin (15-19 Jul 2017),
  \url{http://dx.doi.org/10.1145/3067695.3075965}

\bibitem{langdon:RN1901}
Langdon, W.B., Banzhaf, W.: Faster genetic programming {GPquick} via multicore
  and advanced vector extensions. Tech. Rep. RN/19/01, University College,
  London, London, UK (23 Feb 2019),
  \url{http://www.cs.ucl.ac.uk/fileadmin/user_upload/avx_rn1901.pdf}

\bibitem{Langdon:2019:alife}
Langdon, W.B., Banzhaf, W.: Continuous long-term evolution of genetic
  programming. In: Fuechslin, R. (ed.) Conference on Artificial Life (ALIFE
  2019). pp. 388--395. MIT Press, Newcastle (29 Jul - 2 Aug 2019),
  \url{http://dx.doi.org/10.1162/isal_a_00191}

\bibitem{Handley:1994:DAGpcp}
Handley, S.: On the use of a directed acyclic graph to represent a population
  of computer programs. In: Proceedings of the 1994 IEEE World Congress on
  Computational Intelligence. vol.~1, pp. 154--159. IEEE Press, Orlando,
  Florida, USA (27-29 Jun 1994),
  \url{http://dx.doi.org/10.1109/ICEC.1994.350024}

\bibitem{mcphee:1998:sutherland}
McPhee, N.F., Hopper, N.J., Reierson, M.L.: Sutherland: An extensible
  object-oriented software framework for evolutionary computation. In: Koza,
  J.R., Banzhaf, W., Chellapilla, K., Deb, K., Dorigo, M., Fogel, D.B., Garzon,
  M.H., Goldberg, D.E., Iba, H., Riolo, R. (eds.) Genetic Programming 1998:
  Proceedings of the Third Annual Conference. p. 241. Morgan Kaufmann,
  University of Wisconsin, Madison, Wisconsin, USA (22-25 Jul 1998),
  \url{http://facultypages.morris.umn.edu/~mcphee/Research/Sutherland/sutherland_gp98_announcement.ps.gz}

\bibitem{flajlet:1982:ahbt}
Flajolet, P., Oldyzko, A.: The average height of binary trees and other simple
  trees. Journal of Computer and System Sciences  25(2),  171--213 (October
  1982), \url{https://doi.org/10.1016/0022-0000(82)90004-6}

\bibitem{langdon:2008:eurogp}
Langdon, W.B., Banzhaf, W.: A {SIMD} interpreter for genetic programming on
  {GPU} graphics cards. In: O'Neill, M., Vanneschi, L., Gustafson, S.,
  {Esparcia Alcazar}, A.I., {De Falco}, I., {Della Cioppa}, A., Tarantino, E.
  (eds.) Proceedings of the 11th European Conference on Genetic Programming,
  EuroGP 2008. Lecture Notes in Computer Science, vol. 4971, pp. 73--85.
  Springer, Naples (26-28 Mar 2008),
  \url{http://dx.doi.org/10.1007/978-3-540-78671-9_7}

\bibitem{langdon:2008:SC}
Langdon, W.B., Harrison, A.P.: {GP} on {SPMD} parallel graphics hardware for
  mega bioinformatics data mining. Soft Computing  12(12),  1169--1183 (Oct
  2008), \url{http://dx.doi.org/10.1007/s00500-008-0296-x}, special Issue on
  Distributed Bioinspired Algorithms

\bibitem{langdon:2008:CIGPU}
Langdon, W.B.: A fast high quality pseudo random number generator for graphics
  processing units. In: Wang, J. (ed.) 2008 IEEE World Congress on
  Computational Intelligence. pp. 459--465. IEEE, Hong Kong (1-6 Jun 2008),
  \url{http://www.cs.ucl.ac.uk/staff/W.Langdon/ftp/papers/langdon_2008_CIGPU.pdf}

\bibitem{Park:1988:RNG}
Park, S.K., Miller, K.W.: Random number generators: Good ones are hard to find.
  Communications of the ACM  32(10),  1192--1201 (Oct 1988),
  \url{http://dx.doi.org/10.1145/63039.63042}

\end{thebibliography}

\end{document}